\documentclass[reprint, amsmath,amssymb, superscriptaddress, aps,prb]{revtex4-2}

\usepackage{graphicx}
\usepackage{dcolumn}
\usepackage{bm}
\usepackage{hyperref}
\hypersetup{colorlinks, allcolors=blue}
\usepackage{xcolor}
\usepackage{physics}
\usepackage[version=4]{mhchem} 
\usepackage{siunitx}
\DeclareSIUnit\rydberg{Ry}
\DeclareSIUnit\angstrom{\text{Å}}
\usepackage{tablefootnote}
\usepackage{threeparttable}
\usepackage{soul}

\begin{document}

\preprint{APS/123-QED}

\title{Tailoring the Electronic Structure of Ni(111) by Alloying with Sb Ad-Atoms}

\author{Anna Cecilie {\AA}sland}
\affiliation{Department of Physics, Norwegian University of Science and Technology (NTNU), NO-7491 Trondheim, Norway}

\author{Alv Johan Skarpeid}
\affiliation{Department of Physics and Centre for Materials Science and Nanotechnology, University of Oslo, NO-0318, Oslo, Norway}

\author{Matthias Hartl}
\affiliation{Department of Physics, Norwegian University of Science and Technology (NTNU), NO-7491 Trondheim, Norway}

\author{Marte Stalsberg}
\affiliation{Department of Physics and Centre for Materials Science and Nanotechnology, University of Oslo, NO-0318, Oslo, Norway}

\author{Adrian N. Rørheim}
\affiliation{Department of Physics and Centre for Materials Science and Nanotechnology, University of Oslo, NO-0318, Oslo, Norway}

\author{Johannes Bakkelund}
\affiliation{Department of Physics, Norwegian University of Science and Technology (NTNU), NO-7491 Trondheim, Norway}
\affiliation{Department of Engineering Sciences, University of Agder, NO-4879 Grimstad, Norway}

\author{Jinbang Hu}
\affiliation{Department of Physics, Norwegian University of Science and Technology (NTNU), NO-7491 Trondheim, Norway}

\author{Zheshen Li}
\affiliation{Department of Physics and Astronomy - Centre for Storage Ring Facilities (ISA), Aarhus University, DK-8000 Aarhus, Denmark}

\author{Simon P. Cooil}
\affiliation{Department of Physics and Centre for Materials Science and Nanotechnology, University of Oslo, NO-0318, Oslo, Norway}

\author{Justin W. Wells}
\email[Contact author: ]{quantumwells@gmail.com}
\affiliation{Department of Physics, Norwegian University of Science and Technology (NTNU), NO-7491 Trondheim, Norway}
\affiliation{Department of Physics and Centre for Materials Science and Nanotechnology, University of Oslo, NO-0318, Oslo, Norway}

\author{H{\aa}kon I. R{\o}st}
\email[Contact author: ]{hkonrost@outlook.com}
\affiliation{Department of Physics and Technology, University of Bergen, NO-5007 Bergen, Norway}

\begin{abstract}

Surface alloying can alter surface electronic and magnetic properties, which are key parameters when developing new materials tailored for specific applications.   
A magnetic surface alloy was formed by depositing Sb on Ni(111) at elevated temperatures, yielding new electronic states at the Fermi level and modifying the Ni-derived bandstructure. In particular, it changed the electron occupancy of the spin-polarized surface resonance bands, which may affect the magnetic properties of the surface and its associated many-body effects. By fitting a finite element model to angle-dependent core level measurements, similar amounts of Sb and Ni were found within the first few atomic layers to indicate a near-surface composition similar to the bulk alloy NiSb. Annealing to higher temperatures post-growth further improved the crystalline quality of the surface.
Our investigation of the surface alloy's crystallinity, chemical composition, and layer structure lays the basis for future studies of how its electronic and magnetic properties can be modified.  
\end{abstract}

\maketitle

\section{Introduction}\label{sec:intro}

Many new materials and devices have been developed by combining well-known materials with desired but disparate properties \cite{Palmstroem2003, Hellman2017, Zutic2019, Nonnig2023, Laterza2024}.   
For instance, proximity effects and alloying at the interface between magnetic and non-magnetic materials can modify the magnetism, spin-transport properties and electron-magnon coupling of the involved materials, which can be of importance within spintronics, superconductivity, and the development of magnetic devices \cite{Vasili2018, Zhu2019a, Zutic2019, Hirohata2020, Karchev2003, Rohling2018, Maeland2021, Rost2024disentangling}. 
Strong spin-polarization and spin dynamic effects can also be realized using compounds of magnetic and strong spin-orbit coupled (SOC) materials like NiMnSb \cite{Ciccarelli2016, Gerhard2014, Toual2024}. 
Alloys of Ni and Sb have further applications within thermoelectrics as well as catalysis since the modified surface electronic structure also changes the surface chemistry \cite{Deckers1990, Xu1995, Besenbacher1998, Katsuyama2003, Nikolla2007, Zhang2008, Fan2021}.   

In this work, we explore the composition and properties of crystalline $\mathrm{Ni}_{1-x}\mathrm{Sb}_{x}$ surface alloys on Ni(111). Low-energy electron diffraction (LEED), angle-resolved photoemission spectroscopy (ARPES), and angle-resolved X-ray photoelectron spectroscopy (ARXPS) were used to investigate the crystalline structure, electronic structure, and stoichiometry of the near-surface layers, respectively. The Sb-modified system exhibited a $p(\sqrt{3}\times\sqrt{3})\text{R}30^\circ$ surface periodicity relative to the pristine Ni(111). Additional electronic bands appeared, along with modifications to the pre-existing, Ni-derived bandstructure. Using a finite element approach \cite{song2012extracting,Song2013}, the distribution of Sb into Ni could be traced primarily to the topmost few atomic layers. This surface-localized alloying should strongly affect the magnetization and atomic density of the Ni(111) surface, having important implications for its collective magnetic and non-magnetic excitation states \cite{Bouma2020itinerant,streubel2021chiral}. Therefore, precise alloying with foreign and non-magnetic species such as Sb poses an exciting step towards tailored spin materials with designer-induced properties. In the extreme limit, the surface can host a drastically modified electronic structure and collective excitations \cite{Carrasco:2023}.

\section{Results and Discussion}

\begin{figure}
    \centering
    \includegraphics[]{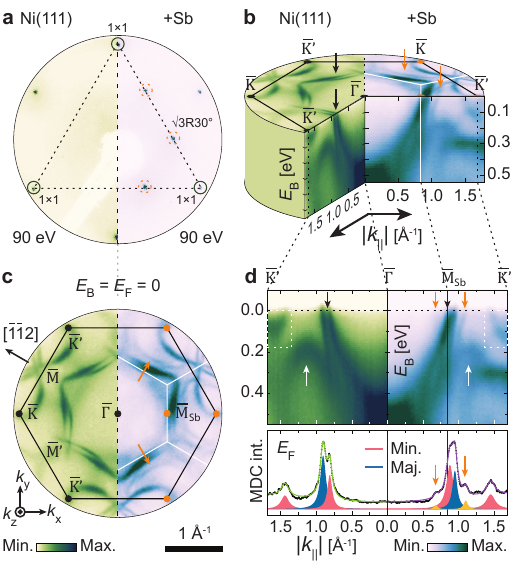}
    \caption{The surface and electronic structure of Ni(111), and Sb-rich layers on the Ni(111) face. \textbf{a}: A split-view of the surface reciprocal lattice without/with additional Sb, as measured using LEED. The 3-fold symmetry of the (111) face is visible from the principal ($1\times1$) spots. The Sb ad-atoms induce a prominent $p(\sqrt{3}\times\sqrt{3})\text{R}30^\circ$ reconstruction. \textbf{b}: Volumetric representation of the measured electronic bandstructure. \textbf{c}: Constant energy surface ($E_{\text{B}}=E_{\text{F}}$) measurements of Ni(111) without/with Sb ad-atoms. The projected bulk Brillouin zone (PBZ, black) and its $p(\sqrt{3}\times\sqrt{3})\text{R}30^\circ$ reconstruction (white) are overlaid, with high-symmetry points indicated. \textbf{d}: $E_{\text{B}}$~vs.~$k_{||}$ measurements (top) of both systems along the $\overline{\Gamma}-\overline{\text{K}}'$ high-symmetry direction of the PBZ. The addition of Sb gives rise to new bands (orange arrows), as seen from the MDC at $E_{\text{B}}=E_{\text{F}}$ (bottom). Additional modifications to the pre-existing Ni bands can also be observed (white arrows/boxes). Spin-band assignments have been based on the theoretical predictions in Refs.~\onlinecite{Osterwalder2006,Rost2024disentangling}.}
    \label{fig:LEEDandARPES}
\end{figure}

Following Krupski \emph{et al.} \cite{KRUPSKI2003}, we prepared alloys of ferromagnetic and strong SOC species by depositing thin films of Sb on Ni(111) surfaces at elevated temperatures (see details in the Supporting Information \cite{Supporting}). When treated to $T>\SI{500}{\kelvin}$, the resultant alloying between Ni and Sb yields a prominent $p(\sqrt{3}\times\sqrt{3})\text{R}30^\circ$ surface structure compared to the pristine Ni(111). This is evidenced by LEED measurements of the surface (Fig.~\ref{fig:LEEDandARPES}\textbf{a}), revealing additional and similarly sharp diffraction spots. Higher growth temperatures and less Sb deposited resulted in sharper diffraction spots on lower-intensity backgrounds, indicating crystalline surfaces with less scattering from defects and impurities \cite{Supporting}.

The observable electronic structures of pristine and Sb-treated Ni(111) surfaces are expected to contain contributions from surface- and bulk-derived energy states \cite{kevan1992angle}. In Figs.~\ref{fig:LEEDandARPES}\textbf{b}-\textbf{d}, momentum microscopy measurements of the two systems are shown side-by-side for comparison \cite{Escher2005NanoESCA,tusche2019imaging}. Along the $\bqty{111}$ direction, the projected bulk Brillouin zone (PBZ) is hexagonal and threefold symmetric (Fig.~\ref{fig:LEEDandARPES}\textbf{b}) \cite{Tserkezis2011photonic}. Within the PBZ of both systems, common and prominent features derived from bulk Ni can be observed near the Fermi level ($E_{\text{F}}$) \cite{Osterwalder2006}. Additionally, the Ni(111) surface resonances (black arrows) are still present after adding Sb \cite{Rost2024disentangling}. Immediately, this suggests a narrow film of mixed Ni and Sb near the surface, or a chemisorbed Sb layer that is intimately in contact with the underlying Ni substrate and interacting with the Ni $3d$ states \cite{Preobrajenski2008adsorption,Usachov2010quasifreestanding}.

Following the addition of Sb, new energy dispersions appear encircling the $\overline{\text{K}}$ and $\overline{\text{K}}'$ high-symmetry points of the PBZ (orange arrows). At the $E_{\text{F}}$, these assume the periodicity of the Sb-induced $p(\sqrt{3}\times\sqrt{3})\text{R}30^\circ$ surface reconstruction (Fig.~\ref{fig:LEEDandARPES}\textbf{c}, white hexagons). Along $\overline{\Gamma}-\overline{\text{K}}'$, they can be readily distinguished from the Ni-derived spin resonances near the zone boundary of the reconstructed surface ($\overline{\text{M}}_{\text{Sb}}$ in Fig.~\ref{fig:LEEDandARPES}\textbf{d}). Alloying with Sb also induces several modifications to the Ni-derived spin bands \cite{Supporting}. Notably, the spin-minority electron `pockets' near the $\overline{\text{K}}'$ and $\overline{\text{K}}$ points (dashed white boxes) shift to larger binding energies $E_{\text{B}}$ by $\approx\SI{40}{\meV}$, and the near-parabolic spin-majority state at $E_{\text{B}}\approx\SI{200}{\meV}$ (white arrows) diminishes. A new band maximum also appears closer to $\overline{\text{K}}'$/$\overline{\text{K}}$ at $E_{\text{B}}\approx\SI{300}{\meV}$. Together, these changes suggest a charge redistribution in the spin-polarized bands that should affect the overall magnetization of the near-surface layers. In turn, this could significantly impact the occurrence, energies, and interaction strengths of the magnetic and non-magnetic many-body effects that are already visible in pure Ni \cite{Rost2024disentangling,Higashiguchi2005energy,Hofmann2009renormalization}.

At present, disentangling and quantifying the effects of Sb alloying -- e.g., on the spin occupancy or the interactions of electrons with collective excitations, requires precise modeling from first principle calculations or lattice-based models \cite{muller2019electron,ke2021electron,nabok2021electron, Hodt2023}. The starting point of any such endeavor would be stable and reproducible configurations, with well-understood compositions exhibiting the altered electronic structure. This task is challenging, as Ni and Sb (in bulk) can form multiple stable phases depending on the stoichiometry and heat treatment involved \cite{Raghavan2004,Okamoto2009}. Confinement to or near the surface further complicates the problem, and the LEED and ARPES (Fig.~\ref{fig:LEEDandARPES}) alone cannot elucidate the detailed chemical composition of the near-surface atomic layers. Hence, we performed high-resolution ARXPS measurements of the Ni~$3p$ and Sb~$4d$ core levels to quantify the amount of Sb, and fitted a layered model of its distribution. 

\begin{figure}
    \centering
    \includegraphics[]{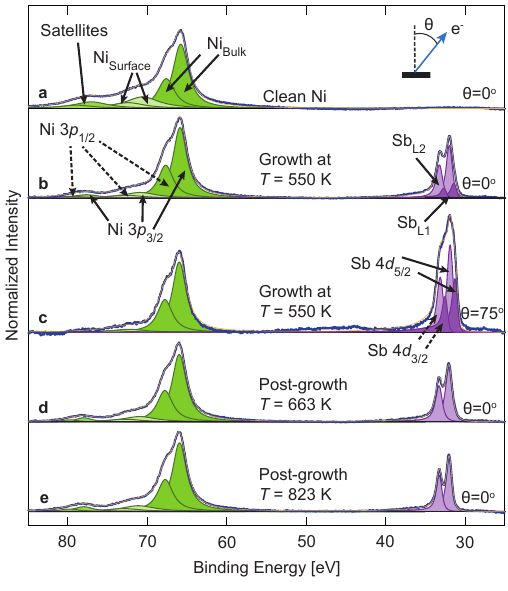}
    \caption{XPS ($h\nu=250$~eV) of the Ni~$3p$ and Sb~$4d$ core levels after different preparation steps. The backgrounds have been subtracted and the spectra normalized to the area of the bulk $\mathrm{Ni}~3\mathit{p}$. \textbf{a}: Pristine Ni(111). \textbf{b}-\textbf{c}: After growth at $T=550$~K. \textbf{d}: After annealing post-growth to $T=663$~K. \textbf{e}:  After annealing post-growth to $T=823$~K. The inset shows the emission angle $\theta$ relative to the sample surface normal.
    }
    \label{fig:NiSb_XPS}
\end{figure}

A synopsis of core-level measurements recorded at different stages of preparation is given in Fig.~\ref{fig:NiSb_XPS}. The various contributions to the Ni~$3p$ and Sb~$4d$ peaks were resolved by fitting pseudo-Voigt functions on a Shirley background \cite{schmid2014new}. To facilitate visual comparison between the different treatments, all spectra have been normalized to the area of the bulk Ni~$3p$ doublet and their backgrounds have been subtracted.

Three doublets were fitted to the $3p$ core level of pristine Ni, each with a spin-splitting of $1.8$~eV (Fig.~\ref{fig:NiSb_XPS}\textbf{a}). 
The component at $E\mathrm{_{B}}=65.8$~eV (dark green) corresponds to metallic Ni in the bulk layers \cite{McIntyre1975, Born2021}. 
The component at $E\mathrm{_{B}}=70.6$~eV (light green) is interpreted as surface Ni, its binding energy similar to that of oxidized $\mathrm{Ni}^{3+}$ \cite{GilMonsalve2021, Qiao2011, Saghayezhian2017}. 
The energy difference between bulk Ni~$3p$ and the satellite peak at $E_{\mathrm{B}}=77.1$~eV (medium green) is roughly similar to an observed energy loss of $11.2$~eV related to the Ni~$3d\rightarrow 4p$ excitation \cite{HagelinWeaver2004}.

With Sb deposition at $T=550$~K (Fig.~\ref{fig:NiSb_XPS}\textbf{b}), the Ni~$3p$ doublets shifts towards larger binding energies by $+0.2$~eV (bulk), $+0.4$~eV (surface), and $+1.0$~eV (satellite). This is presumably due to electron donation to the Sb during alloying~\cite{Fan2021}. Two additional Sb~$4d$ doublets, each with a spin-splitting of $1.3$~eV, emerge at $E_{\mathrm{B}}^{\mathrm{L1}}=31.5$~eV and $E_{\mathrm{B}}^{\mathrm{L2}}=32.1$~eV, respectively. The former matches with metallic Sb and has been assigned accordingly \cite{Morgan1973, Maskery2016, Shiel2019, Fan2021}. The latter has an energy similar to $\mathrm{Sb}^{3+}$ \cite{Fan2021}, but is interpreted as Sb more intimately mixing with the Ni. Complementary ARXPS measurements (Fig.~\ref{fig:NiSb_XPS}\textbf{b}-\textbf{c}) of the alloy as grown revealed both ($i$) a larger relative abundance of $\mathrm{Sb_{L1}}$ vs. $\mathrm{Sb_{L2}}$ at grazing emission, and ($ii$) a larger relative abundance of Sb vs. Ni overall. Immediately, this suggested that the Sb was concentrated in the first few atomic layers, with metallic Sb at the very top \cite{Rost2023probing}. 

To investigate the effect of post-growth-annealing, the $\mathrm{Ni}_{1-x}\mathrm{Sb}_{x}$ alloy was heated in steps to increasingly higher temperatures up to \SI{1130}{\kelvin}. Consequently, the Sb either evaporated or diffused into the underlying bulk ferromagnet \cite{Rost2021low}, as evidenced by the relative abundance of Sb~$4d$ vs. Ni~$3p$ in Figs.~\ref{fig:NiSb_XPS}\textbf{d}-\textbf{e}. Furthermore, the $\mathrm{Sb_{\mathrm{L1}}}$ signal weakened while the $\mathrm{Sb_{L2}}$ signal remained virtually unchanged. This change is consistent with the $\mathrm{Sb_{L2}}$ bonding more strongly to the Ni.

\begin{figure}
    \centering
    \includegraphics[width=1.0\linewidth]{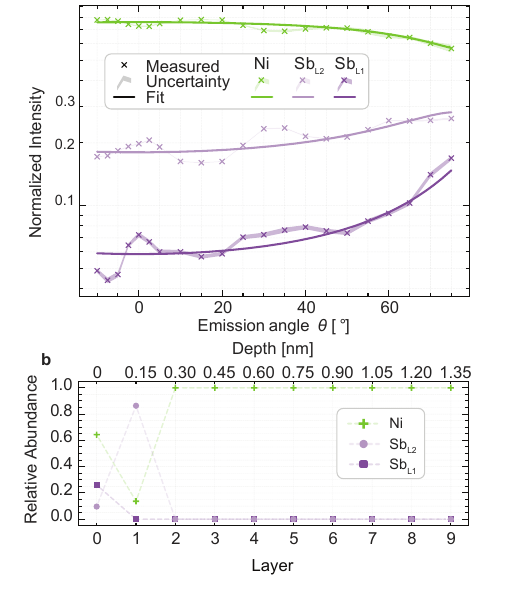}
    \caption{The finite element model with $10$ layers and a ratio $\dd z / \lambda = 0.3$ applied to a sample grown at $T=550$~K. \textbf{a:} The fitted normalized intensity on a logarithmic scale as a function of emission angle $\theta$ for {Ni} and the {Sb}$_{\text{L1}}$ and {Sb}$_{\text{L2}}$ components. The associated $\chi^2$ value of the fit is $\chi^2_{(1)}/N = 86.2$. \textbf{b:} The corresponding depth profile (markers) joined by dashed lines. The fitted composition suggests the Sb is confined to the near-surface layers, with the Sb$_{\text{L1}}$ only at the very surface.
    }
    \label{fig:NiSb_depth}
\end{figure}

To examine the layer-by-layer composition and stoichiometry of the surface alloy as grown (i.e, before the post-growth-annealing), additional ARXPS measurements were acquired at emission angles between $\theta=-10^{\circ}$ and $\theta=+75^{\circ}$. The results were fitted to a simple finite element layer model following the approach of Refs.~\onlinecite{song2012extracting, Song2013}. In the model, each layer is of equal finite thickness \(\dd z\) and contains a relative amount \(Q^{(\text{A})}\) of each of the chemical species \((\text{A})\) measured. The key model assumption is that the signal intensity from each species \(I^{(\text{A})}\) depends on the emission angle \(\theta\) (see the inset in Fig.~\ref{fig:NiSb_XPS}) as
\begin{equation}\label{eq:FEModel}
    I^{(\text{A})} \propto
    \sum_{i} Q^{(\text{A})}_{i}
    \sigma^{(\text{A})} \exp\left(- \, \frac{z_i}{\lambda\cos{\theta}}\right),
\end{equation}
where, \(i\) is the index of the layers; \(z_i\geq 0\) is the depth to the top of the \(i\)-th layer from the surface and layer \(i = 0\) is at zero depth; \(\lambda\) is the inelastic mean free path; \(\sigma^{(\text{A})}\) is the photoionization cross-section of specie \(\text{A}\); and \( Q^{(\text{A})}_{i}\) is the proportion of \(\text{A}\) in layer \(i\) normalized such that \(\sum_A Q^{(A)}_{i} = 1\). The attenuation length \(\lambda\) was assumed to depend only on the photoelectron kinetic energy ($E_{\mathrm{K}}$), and the photoionization cross-sections \(\sigma^{(\text{A})}\) were obtained from Refs.~\onlinecite{Yeh1985atomic,yeh1993atomic}. Further details regarding the implementation have been given in the Supporting Information~\cite{Supporting}. 

The results from the finite element model analysis are shown in Fig.~\ref{fig:NiSb_depth}, with uncertainties taken from the deconvolution analysis of the individual core levels. The fit reproduces the overall angle dependence of the measured core levels. However, the achievable goodness of fit metric is $\chi^{2}/N\approx 86$, suggesting that the recorded uncertainties are small compared to the overall fit residuals. The large $\chi^2$ value should be interpreted as the unmodeled photoelectron diffraction effects \cite{Woodruff1994, Woodruff2007} being much stronger than the measured signal-to-noise ratio. The photoelectron diffraction manifests as oscillations in the angle-dependent (normalized) intensities. These originate primarily in the raw Ni signal \cite{Fadley1987photoelectron,Sambi1995photoelectron,Parreiras2014graphene}, and propagate to the Sb signal as a consequence of the intensity normalization.

We tuned the model hyperparameters by minimizing the \(\chi^2\)-metric using grid search and fitting it to the entire dataset \cite{nocedal1999numerical, hastie01statisticallearning}. With a sufficient number of layers, the optimal ratio between layer thickness and inelastic mean free path was $\dd z/\lambda\approx0.3$. Using different hyperparameters, the model yielded slightly different results but with a similar goodness of fit. Assuming $\lambda\approx 0.5$~nm for photoelectrons with $E_{\mathrm{K}}=170$--$220$~eV \cite{seah1979quantitative,Seah1980,Zangwill1988,Powell2020}, \(\dd z\approx\SI{0.15}{\nm}\) and comparable to the thickness of a single atomic layer.

The fitted depth profile in Fig.~\ref{fig:NiSb_depth}\textbf{b} suggests a $\mathrm{Ni}\mathrm{Sb}$ alloy is formed within the first two model layers with total thickness $t=\SI{0.3}{\nm}$. The metallic Sb from $\mathrm{Sb_{L1}}$ can be found closest to the surface. As the $\mathrm{Sb_{L1}}$ component disappears at higher temperatures (Fig.~\ref{fig:NiSb_XPS}), we interpret it as less stable and at most weakly alloying with the Ni. By treating the topmost two layers as a single slab of some alloy, the ratio of $\mathrm{Sb_{L2}}$ to Ni is $1.2:1$ ($x\approx 0.55$). We thus interpret $\mathrm{Sb_{L2}}$ as Sb atoms mixing with Ni near the surface to form near-stoichiometric NiSb \cite{Okamoto2009,Jain2013,Nisb}.

At $T>\SI{800}{\K}$, the relative strength of the $\mathrm{Sb}$ signal reduced by $\approx 1/3$ indicating that either Sb was lost from the sample, diffused further into the bulk Ni, or both. The data and our modeling support the notion of the alloy region broadening without significant loss of $\mathrm{Sb}$ \cite{Supporting}. At even higher temperatures the $\mathrm{Sb_{L2}}$ signal remained virtually unchanged, consistent with an Sb-deficient phase of NiSb that is stable up to \SI{1350}{\K} \cite{Okamoto2009}.

Finally, we deposited larger amounts of Sb during growth to investigate the diffusion of Sb into the bulk Ni and potentially thicker alloy films. The ARXPS and associated finite element modeling suggested a thicker alloy region with more metallic Sb on the surface as a result \cite{Supporting}. Bulk-like NiSb alloys should exhibit different electronic and magnetic properties than the NiSb surface alloys already discussed (Figs.~\ref{fig:LEEDandARPES}-\ref{fig:NiSb_depth}). We conjecture that tailoring the properties of NiSb and its sister phases by transitioning between surface and bulk alloys can be an exciting avenue for further investigations.

\section{Conclusion}
Depositing atomic Sb onto Ni(111) at elevated temperatures forms a crystalline and Sb-rich surface alloy with newfound and altered electronic states. The occupancy of the Ni-derived, spin-polarized bands changes, potentially influencing the collective excitation states and associated many-body interactions of the surface. The chemical composition and thickness of the surface alloy were investigated using ARXPS and finite element layer modeling, indicating similar amounts of Sb and Ni in a $t\approx\SI{0.3}{\nm}$ thick layer. Additional traces of metallic Sb could also be observed. Annealing to higher temperatures diminished the metallic Sb but did not significantly alter the composition of the stable NiSb phase. Our study of the stoichiometry, crystallinity, and electronic structure of the NiSb surface alloy can offer valuable insights for concurrent investigations of the collective excitation states of alloyed ferromagnets.

\section{Acknowledgements}

This work was partly supported by the Research Council of Norway, project numbers 324~183, 315~330,  335~022, and 262~633. We acknowledge funding via the ‘Sustainable Development Initiative’ at UiO. The authors acknowledge the staff at the ASTRID2 synchrotron in Aarhus, Denmark, for providing access to their synchrotron radiation facilities, and for practical assistance and discussions. We also thank C. J. Palmstrøm and F. Hellman for insightful discussions on magnetic alloying.

%

\end{document}


\title{Supporting Information: Tailoring the Electronic Structure of Ni(111) by Alloying with Sb Ad-Atoms}

\author{Anna Cecilie {\AA}sland}
\affiliation{\footnotesize{Department of Physics, Norwegian University of Science and Technology (NTNU), NO-7491 Trondheim, Norway}}

\author{Alv Johan Skarpeid}
\affiliation{\footnotesize{Department of Physics and Centre for Materials Science and Nanotechnology, University of Oslo, NO-0318, Oslo, Norway}}

\author{Matthias Hartl}
\affiliation{\footnotesize{Department of Physics, Norwegian University of Science and Technology (NTNU), NO-7491 Trondheim, Norway}}

\author{Marte Stalsberg}
\affiliation{\footnotesize{Department of Physics and Centre for Materials Science and Nanotechnology, University of Oslo, NO-0318, Oslo, Norway}}

\author{Adrian N. Rørheim}
\affiliation{\footnotesize{Department of Physics and Centre for Materials Science and Nanotechnology, University of Oslo, NO-0318, Oslo, Norway}}

\author{Johannes Bakkelund}
\affiliation{\footnotesize{Department of Physics, Norwegian University of Science and Technology (NTNU), NO-7491 Trondheim, Norway}}
\affiliation{\footnotesize{Department of Engineering Sciences, University of Agder, NO-4879 Grimstad, Norway}}

\author{Jinbang Hu}
\affiliation{\footnotesize{Department of Physics, Norwegian University of Science and Technology (NTNU), NO-7491 Trondheim, Norway}}

\author{Zheshen Li}
\affiliation{\footnotesize{Department of Physics and Astronomy - Centre for Storage Ring Facilities (ISA), Aarhus University, DK-8000 Aarhus, Denmark}}

\author{Simon P. Cooil}
\affiliation{\footnotesize{Department of Physics and Centre for Materials Science and Nanotechnology, University of Oslo, NO-0318, Oslo, Norway}}

\author{Justin W. Wells}
\email[Contact author: ]{quantumwells@gmail.com}
\affiliation{\footnotesize{Department of Physics, Norwegian University of Science and Technology (NTNU), NO-7491 Trondheim, Norway}}
\affiliation{\footnotesize{Department of Physics and Centre for Materials Science and Nanotechnology, University of Oslo, NO-0318, Oslo, Norway}}

\author{H{\aa}kon I. R{\o}st}
\email[Contact author: ]{hkonrost@outlook.com}
\affiliation{\footnotesize{Department of Physics and Technology, University of Bergen, NO-5007 Bergen, Norway}}

\maketitle

\section{Experimental Details}

\subsection{Sample Preparation}\label{sec:samplePrep}

The Ni(111) surface was cleaned in ultrahigh vacuum by cycles of Ar$^+$ ion sputtering at $800-1000$~eV and subsequent annealing to $T\approx873$~K. The cleanliness and crystallinity of the surface were verified using X-ray photoelectron spectroscopy (XPS) and low-energy electron diffraction (LEED), respectively. The threefold symmetric surface structure of clean Ni(111) is evident from the diffraction pattern shown in the left-hand-side panel of Fig.~\ref{fig:structure}\textbf{a}.

Alloys of Ni and Sb were prepared by thermally evaporating atomic Sb from a Ta crucible onto a heated Ni(111) surface. The Ni(111) was kept at $500$~--~$700$~K during growth, and the Sb deposition terminated after achieving a $\approx3:1$ ratio between the cross-section-normalized Ni~3\textit{p} and Sb~4\textit{d} core levels (Sb thickness $<0.5$~nm as found using a primitive two-layer model \cite{Rost2021low}). Subsequent annealing (post-growth) was performed in steps between $600$~--~$800$~K to recrystallize the surface ($10$~--~$15$~min/step). The mixing of Sb and Ni at the surface yielded crystalline alloy layers with a relative $p(\sqrt{3}\times\sqrt{3})\text{R}30^\circ$ superstructure as compared to pristine Ni(111) (see Sec.~\ref{sec:LEED} for additional details). 

Heating during and post-Sb deposition was mainly achieved by bombarding thermionic electrons from a nearby hot filament. The resulting sample temperatures were precisely measured with a thermocouple mounted on the sample plate, close to the Ni crystal. Table~\ref{tab:SamplePrepTC} summarizes the preparation parameters and lists any corresponding XPS figures in the text.

Any additional NiSb alloys (Fig.~\ref{fig:LEED_prep}) were heated radiatively during and after growth, using an adjacent filament. In this case, the sample temperatures were estimated based on the filament power $P \propto T^{4}$ (Table~\ref{tab:SamplePrepLEED})~\footnote{Assuming $T=\alpha \times P^{1/4} + \SI{300}{\kelvin}$ for the sample temperature. The proportionality constant $\alpha$ was determined using a thermocouple mounted onto a solid copper piece in contact with the sample holder that acted as a thermal conductor for the sample manipulator.}. These values represent a minimum estimate of the effective sample temperatures. However, by comparing the resulting LEED (Fig.~\ref{fig:LEED_prep}), and Ni~3\textit{p} and Sb~4\textit{d} core levels to those achieved using the growth parameters in Table~\ref{tab:SamplePrepTC}, we could crudely estimate the true growth and post-growth temperatures with filament heating. These corrected temperature estimates have been listed in the `Corrected' columns of Table~\ref{tab:SamplePrepLEED}. The same table also lists the optimized sample recipe, which yields the bandstructure shown in Figs.~1 (main text) and \ref{fig:SI_ARPES}.

\begin{table}[t]
    \centering
    \caption{Sample preparation temperatures during and post-growth for surface alloys of Ni and Sb. All temperatures were measured using a K-style thermocouple on the sample plate and close to the Ni(111) substrate. Figures with corresponding XPS core levels have also been indicated.}
    \vspace{0.35cm}
    \begin{tabular}{lccc}
        \hline
        Sample & Growth $T$ [K] & $\quad$Post-growth~$T$ [K] & $\quad$Post-growth anneal time / step [min] \\ \hline
        Figs.~2, 3 & $\quad550$ & $\quad663$~--~$1130$ & $\quad\quad10$ \\ 
        Figs.~\ref{fig:Supp_XPS}, \ref{fig:SI_ARXPS} & $\quad523$ & $\quad\quad623$ & $\quad\quad15$ \\ 
        \hline
    \end{tabular}
    \label{tab:SamplePrepTC}
\end{table}

\begin{table}
    \centering
    \caption{Sample preparation parameters for Ni + Sb surface alloys, prepared using a radiative heating source (filament). The corresponding LEED pattern for each preparation is displayed in Fig.~\ref{fig:LEED_prep}. Growth parameters for the optimized NiSb sample recipe (Fig.~1\textbf{a} in the main text) have been highlighted (green/shaded row). The temperatures in columns `$T\propto P^{1/4}$' were estimated from the operating power of a sample-adjacent filament. The `Corrected' temperatures were estimated by comparing the LEED and XPS results to those achieved from the recipes in Table~\ref{tab:SamplePrepTC}.}
    \vspace{0.35cm}
    \begin{tabular}{lcccccc}
        \hline
        Sample & Sb deposition$\quad$ & \multicolumn{2}{c}{Growth $T$ [K]} & \multicolumn{2}{c}{Post-growth~$T$ [K]}$\quad\quad$ & Post-growth anneal \\
         & time [min]$\quad$ & $T\propto P^{1/4}$ & Corrected$\quad$  & $T\propto P^{1/4}$ & Corrected$\quad\quad$ & time [min] \\ \hline 
        Fig.~\ref{fig:LEED_prep}\textbf{a} & 5$\quad$ & 500 & 670$\quad$  & 600 & 800$\quad\quad$ & 12 \\
        Fig.~\ref{fig:LEED_prep}\textbf{b} & 5$\quad$ & 500 & 670$\quad$  & 600 & 800$\quad\quad$ & 15 \\
        Fig.~\ref{fig:LEED_prep}\textbf{c} & 5$\quad$ & 530 & 700$\quad$  & 600 & 800$\quad\quad$ & 15 \\
        Fig.~\ref{fig:LEED_prep}\textbf{d} & 3$\quad$ & 510 & 690$\quad$  & 600 & 800$\quad\quad$ & 15 \\
        \rowcolor{lime} Fig.~\ref{fig:LEED_prep}\textbf{e}$\quad$ & 5$\quad$ & 510 & 690$\quad$  & 600 & 800$\quad\quad$ & 15 \\
        Fig.~\ref{fig:LEED_prep}\textbf{f} & 10$\quad$ & 510 & 690$\quad$  & 600 & 800$\quad\quad$ & 15 \\
        \hline 
    \end{tabular}
    \label{tab:SamplePrepLEED}
\end{table}

\subsection{Angle-Resolved Core Level Measurements}

Angle-resolved X-ray photoelectron spectroscopy (ARXPS) was measured using a SPECS PHOIBOS 150 electron analyzer at the AU-MatLine beamline on the ASTRID2 synchrotron facility at Aarhus University. Measurements were performed with a photoexcitation energy of $h\nu=250$~eV and a pass energy $E_{\text{P}}=20$~eV. The energy resolution was in the range of $0.2$~--~$0.3$~eV. The pressure in the chamber was kept in the order of $10^{-10}$~mbar during all the measurements. 

\subsection{Electronic Bandstructure Measurements}\label{sec:expARPES}
The electronic band structure of Ni(111) -- with and without added Sb, was measured using an aberration-corrected and energy-filtered photoemission electron microscope (EF-PEEM) instrument (NanoESCA III, FOCUS GmbH) \cite{Escher2005NanoESCA}. All samples were electrically grounded while subjected to He~I photoexcitation ($h\nu=21.22$~eV), and the microscopy lens column was biased at 12~kV to extract and capture the photoelectrons. Using a  $\leq0.5$~mm entrance slit to the energy filter and a photoelectron pass energy of $E_{\text{P}}=25$~eV, the achievable energy and momentum resolutions were better than $\Delta E = 50$~meV and $\Delta k = 0.02~\text{Å}^{-1}$, respectively. The electronic band structures were reconstructed from constant energy surface (CES) momentum micrographs, i.e., measurements of $E(k_{x},k_{y})$ in the diffractive plane of the instrument \cite{tusche2019imaging}. Each CES was acquired with an effective $k$-space field of view of $3.6~\text{Å}^{-1}$ and with a relative energy step of $+\SI{10}{\milli\eV}$. All measurements were restricted to surface regions of $\approx60\times\SI{60}{\um}^{2}$ using an iris aperture, and were performed at $T\approx\SI{115}{\kelvin}$. The base pressure in the chamber was in the order of $10^{-10}$~mbar. 

\section{Low Energy Electron Diffraction}
\label{sec:LEED}

\subsection{Determining the Coordination of Ni and Sb Atoms at the Surface}

\begin{figure}
    \centering
    \includegraphics[width=\linewidth]{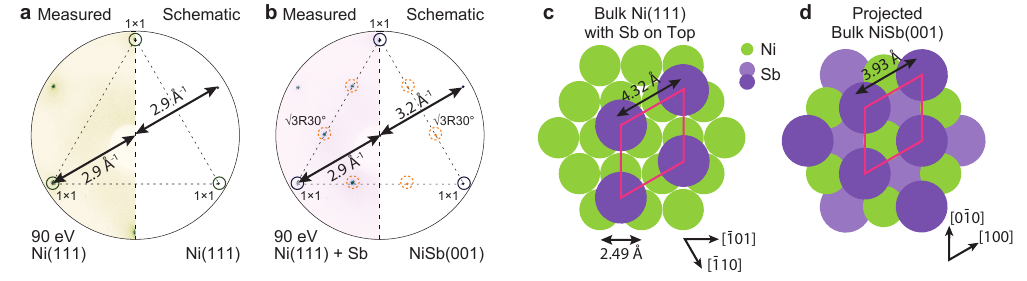}
    \caption{\textbf{a}-\textbf{b}: Schematic of LEED patterns of Ni(111) (\textbf{a}) and NiSb(001) (\textbf{b}). The schematic shows the idealized LEED pattern scaled to the size of the measurement. The sketched $\sqrt{3}$R$30^{\circ}$ spots are placed along the dashed lines, while the dashed circles are placed around the measured spots and reflected about the vertical dashed line. \textbf{c}-\textbf{d}: Top view of the crystal structure of (\textbf{c}) Ni(111) with Sb atoms on the surface and (\textbf{d}) bulk NiSb(001). Some of the primitive bulk unit cell base vector directions are indicated. The configuration in \textbf{c} is based on the stacking suggested in Ref.~\onlinecite{KRUPSKI2003}. Without the Sb ad-atoms, this would be pristine Ni(111) and give rise to a similar LEED pattern as shown in \textbf{a}. Both the ad-atom configuration (\textbf{c}) and the NiSb(001) crystal structure (\textbf{d}) should yield similar diffraction results to the measured LEED pattern in \textbf{b}.}
    \label{fig:structure}
\end{figure}

The crystallinity of the Ni(111) surface alloy before and after the addition of Sb was investigated using LEED (Fig.~\ref{fig:structure}). Adding Sb to Ni(111) at temperatures $T\geq\SI{500}{\kelvin}$ gave rise to additional diffraction spots that indicated a newfound $p(\sqrt{3}\times\sqrt{3})\text{R}30^\circ$ superstructure in the near-surface layers (Fig.~\ref{fig:structure}\textbf{b}, dashed circles).

Krupski \emph{et al.} \cite{KRUPSKI2003} previously observed the layer-by-layer growth of Sb on Ni(111). At $T<\SI{300}{\kelvin}$, small amounts of Sb arranged in a $p(2\times2)$ superstructure, but this transformed into a $p(\sqrt{3}\times\sqrt{3})\text{R}30^\circ$ when heated above $\SI{500}{\kelvin}$. The authors proposed a simple two-layer model of Sb on Ni(111) similar to Fig.~\ref{fig:structure}\textbf{c}. However, complementary Auger electron spectroscopy (AES) measurements indicated that Ni and Sb began to alloy at $T>\SI{300}{\kelvin}$, limiting the achievable concentration of Sb in the near-surface layers. Consequently, it is unclear if the observable $p(\sqrt{3}\times\sqrt{3})\text{R}30^\circ$ structure was solely due to surface Sb ad-atoms (Fig.~\ref{fig:structure}\textbf{c}), a crystalline surface alloy of Ni and Sb, or a combination of both. Given an ordered alloy phase with the right coordination geometry and stoichiometry, the diffraction peaks (Fig.~\ref{fig:structure}\textbf{b}) could well have resulted from the mixing of Ni and Sb near the surface.

The (001) surface-projected crystal structure of bulk NiSb is shown in Fig.~\ref{fig:structure}\textbf{d} \cite{Jain2013, Nisb}. NiSb has a hexagonal unit cell with a similar lattice constant ($\delta a<10\%$) to the surface unit cell of the previously suggested Sb-on-Ni superstructure (Fig.~\ref{fig:structure}\textbf{c}). However, contrary to the Sb-deficient ($1:3$) ad-layer configuration, NiSb is stoichiometrically balanced ($1:1$) with Ni and Sb in separate atomic layers. Our ARXPS measurements and concomitant finite element modeling (Sec.~\ref{sec:lmodel}) yielded a $1.2:1$ ratio between Sb and Ni in the topmost few atomic layers, and diminishing amounts of Sb dispersed in the Ni bulk (see the main text). Thus when combined, our findings indicate that a strained surface alloy of NiSb(001) was formed on top of bulk Ni(111). While other phases of Ni and Sb were also considered, they could be ruled out due to their different stoichiometries and symmetries \cite{Raghavan2004,Okamoto2009}. 

\subsection{Sample Growth Optimization}

\begin{figure}[t]
    \centering
    \includegraphics[]{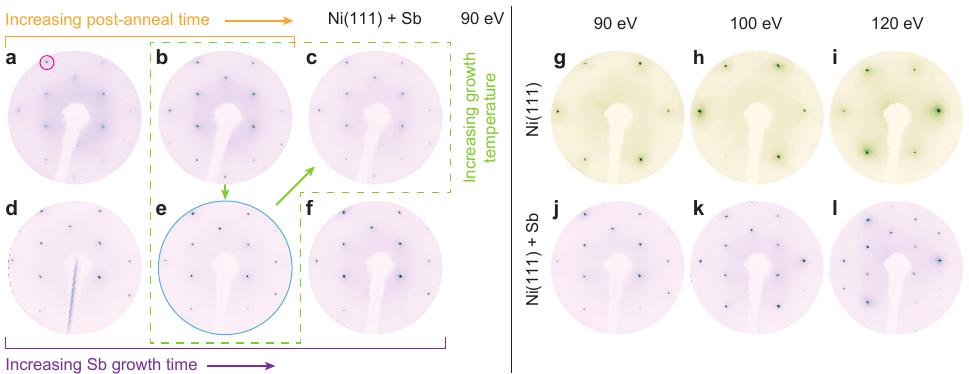}
    \caption{Surface diffraction measurements of Ni(111) with/without added Sb. \textbf{a}-\textbf{f}: Resultant LEED of Ni + Sb for various sample preparation recipes (details in Table~\ref{tab:SamplePrepLEED}). \textbf{g}-\textbf{i}: LEED of pristine Ni(111) with different incident electron energies.
    \textbf{j}-\textbf{l}: LEED of the NiSb surface alloy, prepared using the optimized sample recipe in Table~\ref{tab:SamplePrepLEED} (same as in \textbf{e}). All images have been normalized to the intensity of the left uppermost spots (small circle in \textbf{a}).}
    \label{fig:LEED_prep}
\end{figure}

Figs.~\ref{fig:LEED_prep}\textbf{a}-\textbf{f} contain LEED data from growths of surface NiSb where the post-growth anneal time, growth temperature, and Sb deposition time have been varied (see details in Table~\ref{tab:SamplePrepLEED}). All LEED pattern intensities were normalized to the intensity of the brightest diffraction spot at 90~eV (top left, encircled in Fig.~\ref{fig:LEED_prep}\textbf{a}).
Any observable intensity variation between the symmetry-equivalent spots in the pattern was likely due to the misalignment of the sample plane normal relative to the electron source. 

As evident from Figs.~\ref{fig:LEED_prep}\textbf{a} and \ref{fig:LEED_prep}\textbf{b}, increasing the post-growth anneal times improves the overall crystallinity of the surface. On the other hand, lowering the growth temperature and/or increasing the Sb deposition time roughens the surface, as indicated by the increasing background intensity between Figs.~\ref{fig:LEED_prep}\textbf{b} and \ref{fig:LEED_prep}\textbf{e}, and from  Figs.~\ref{fig:LEED_prep}\textbf{d}-\textbf{f}, respectively. Balancing these three parameters yields an optimized surface preparation as highlighted in Fig.~\ref{fig:LEED_prep}\textbf{e} and detailed in Table~\ref{tab:SamplePrepLEED}. 

\section{Bandstructure Analysis}

\begin{figure}[t]
    \centering
    \includegraphics[]{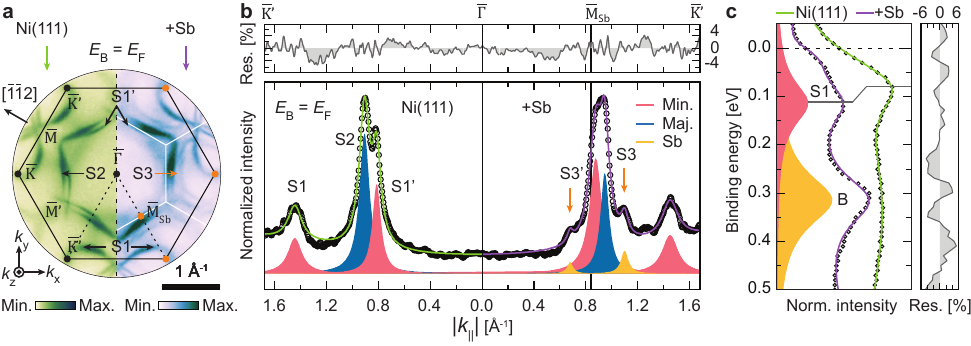}
    \caption{Complementary analysis of the ARPES measurements in Fig.~1 (main text). \textbf{a}: Fermi surface ($E_{\text{B}}=E_{\text{F}}$) measurements of Ni(111) without/with Sb ad-atoms. The projected bulk Brillouin zone (PBZ, black) and its $p(\sqrt{3}\times\sqrt{3})\text{R}30^\circ$ reconstruction (white) have been overlaid with high-symmetry points indicated. \textbf{b}: Fitted MDCs of the pristine (left) and Sb-treated Ni(111) surface (right) along the $\overline{\Gamma}-\overline{\text{K}}'$ high-symmetry line of the PBZ. With Sb added, new states ($\text{S}_{3}^{'}$, $\text{S}_{3}$) can be distinguished near $\overline{\text{M}}_{\text{Sb}}$, assuming the symmetry of the reconstructed surface. \textbf{c}: Fitted EDCs at the local band minimum of the $\text{S}_{1}$ electron `pockets' ($|k_{||}|=1.62~\text{\AA}^{-1}$) along $\overline{\Gamma}-\overline{\text{K}}'$, before and after Sb treatment. All fitted peaks have been approximated using symmetric Lorentzian line shapes, and the intensity backgrounds using 2$^\text{nd}$ degree polynomials. The intensity step at the $E_{\text{F}}$ in \textbf{c} was modeled using a thermally broadened Fermi-Dirac function ($T=\SI{115}{\kelvin}$).}
    \label{fig:SI_ARPES}
\end{figure}

To certify any changes to the measured bandstructure of Ni(111) caused by the addition of Sb, the bandstructure near the Fermi level ($E_{\text{F}}$) was measured as detailed in Sec.~\ref{sec:expARPES}. Constant Fermi energy surfaces of pristine Ni(111) and with Sb ad-atoms have been shown side-by-side in Fig.~\ref{fig:SI_ARPES}\textbf{a}. Therein, well-known spin-polarized states ($\text{S}_1$, $\text{S}_{1}^{'}$, $\text{S}_2$) derived from the Ni bandstructure have been labeled \cite{Osterwalder2006,Rost2024disentangling}. The $\text{S}_{1}^{'}$ and $\text{S}_{1}$ have been interpreted as the same spin-minority band crossing the $E_{\text{F}}$ at different $\vb{k}$ points on the Fermi contour \cite{Rost2024disentangling}. 

Upon alloying with Sb new energy states ($\text{S}_{3}^{'}$, $\text{S}_{3}$) form and assume the periodicity of the $p(\sqrt{3}\times\sqrt{3})\text{R}30^\circ$ surface superstructure (see the LEED in Fig.~\ref{fig:LEED_prep}). As revealed from the fitted momentum distribution curves (MDCs) in Fig.~\ref{fig:SI_ARPES}\textbf{b}, the newfound states appear on either side of the Ni-derived spin-resonance states $(\text{S}_{1}^{'},\text{S}_{2})$ at an approximately similar distance $|\Delta k_{||}|$ to the $\overline{\text{M}}_{\text{Sb}}$ point. They have thus been interpreted as the same energy states, but visible in the primary and next PBZ of the reconstructed surface. 

The added Sb and associated alloying also seem to modify the occupancy of the Ni-derived, spin-polarized energy states. In Fig.~\ref{fig:SI_ARPES}\textbf{c}, energy distribution curves (EDCs) have been extracted at the local band minimum of the $\text{S}_{1}$ spin-minority electron `pockets' along $\overline{\Gamma}-\overline{\text{K}}'$ and near the PBZ edge ($|k_{||}|=1.62~\text{\AA}^{-1}$). With Sb added, the minimum of the spin-minority pocket shifts to larger binding energy by $\approx40$~meV, and a new and prominent feature B appears. Although its origin is not known, the redistribution of photoemission intensity away from its adjacent spin-majority band (white arrows in Fig.~4, main text) may suggest that B is spin-polarized and of majority character \cite{Rost2024disentangling}. The newfound states $(\text{S}_{3}^{'},\text{S}_{3},\text{B})$ and associated modifications to the Ni-derived spin bands thus indicate a changing magnetic occupancy within the near-surface layers. This, in turn, can affect the presence and strength of the boson-derived many-body effects associated with the Ni surface \cite{Higashiguchi2005energy,Hofmann2009renormalization,Rost2024disentangling}.

\section{Temperature-Dependent XPS}
\label{sec:TempDepXPS}
\begin{figure}
    \centering
    \includegraphics[width=0.9\textwidth]{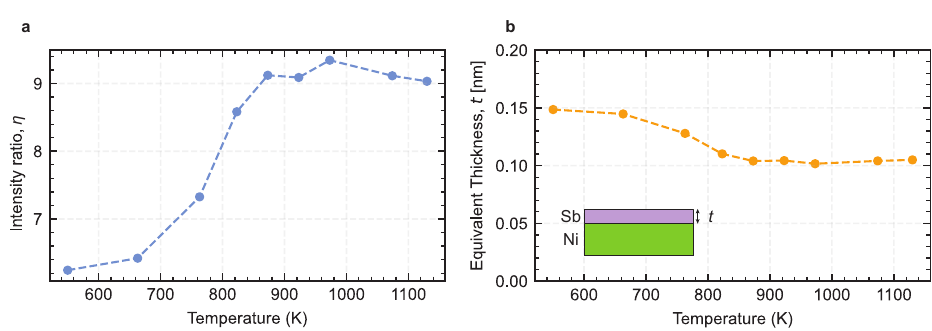}
    \caption{Ratio of the Ni~$3p$ and Sb~$4d$ photoemission intensities in a Ni + Sb surface alloy, as a function of post-growth anneal temperature, corrected for any differences in photoionization cross-section. \textbf{a:} The unprocessed intensity ratio $\eta = I_{\mathrm{Ni}}\sigma_{\mathrm{Sb}}/I_{\mathrm{Sb}}\sigma_{\mathrm{Ni}}$ vs. post-growth anneal temperature. \textbf{b:} The intensity ratio converted to the equivalent thickness $t$ of an Sb layer covering a bulk Ni crystal. The inelastic mean free path is assumed to be $\lambda = \SI{1}{\nm}$ for {$E_{\mathrm{K}} \approx \SI{580}{\eV}$.}}
    \label{fig:temp_series}
\end{figure}

To assess the thermal stability of the NiSb surface alloy, the photoemission intensity ratio of Ni~$3p$~vs.~Sb~$4d$ was recorded as a function of increasing post-growth annealing temperature. The results have been plotted in Fig~\ref{fig:temp_series}\textbf{a}, normalized to their photoionization cross-sections. As a proxy for the abundance of Sb after each post-growth annealing treatment, a simple 2-layer model of Sb placed on top of Ni with Beer-Lambert-like attenuation of the XPS signals was assumed \cite{seah1979quantitative, Seah1980, Powell2020}. 
Therein, an equivalent film thickness $t$ of surface Sb can be calculated as
\begin{equation}
\label{eq:BeerLambert}
    \frac{t}{\lambda(E_{\text{K}})} = \ln \left(
    {1/\eta} + 1
    \right), \quad\quad {\eta = \frac{I_{\mathrm{Ni}}\sigma_{\mathrm{Sb}}}{I_{\mathrm{Sb}}\sigma_{\mathrm{Ni}}}},
\end{equation}
with $\lambda$ being the inelastic mean free path of photoelectrons with kinetic energy $E_{\mathrm{K}}$, and ${I}_\mathrm{A}$ and $\sigma_{\mathrm{A}}$ the photoemission intensity and photoionization cross-section of chemical species A, respectively. Despite its simplifications -- especially for alloys, the estimated equivalent film thickness $t$ is an instructive measure of the relative abundance of Sb. Effectively, $t$ will inform on the Sb concentration over the length scale of the photoelectron inelastic mean free path \cite{Rost2023probing}. Thus for similar $\lambda$, a larger $t$ translates to a higher concentration of Sb in the near-surface atomic layers. The equivalent Sb thicknesses $t$ following post-growth annealing between $T=550$--\SI{1130}{\kelvin} have been plotted in Fig.~\ref{fig:temp_series}\textbf{b}. 

When heated to subsequently higher temperatures between 700 and $\SI{900}{\kelvin}$ post-growth, the relative intensity of Ni~$3p$ signal increased. This translates to a reduced equivalent thickness $t$ by a factor of $\approx1/3$. 
In this temperature range, any loss of observable Sb can mainly be assigned to dilution of the Sb as well as diffusion further into the bulk Ni \cite{KRUPSKI2003}. Thus, post-growth annealing beyond $\SI{700}{\kelvin}$ broadens the alloy region in the direction perpendicular to the surface. At temperatures larger than $\SI{900}{\kelvin}$, however, the equivalent thickness (and Sb concentration) remains similar. This is consistent with an Sb-deficient phase of bulk NiSb that is stable up to $\SI{1350}{\kelvin}$ \cite{Okamoto2009}.

\section{Finite element layer model}
\label{sec:lmodel}
To determine the stoichiometry and depth profile of the prepared NiSb alloys, we fitted the measured ARXPS spectra using a simple finite element layer model, following the approach described in Refs.~\onlinecite{song2012extracting, Song2013}. Each layer in this model is of equal finite thickness \(\dd z\) and contains a relative amount \(Q^{(\text{A})}\) of each of the chemical species \((\text{A})\) measured. The key model assumption is that the signal intensity from each species \(I^{(\text{A})}\) depends on the photoemission angle \(\theta\) relative to the surface normal as
\begin{equation}\label{eq:FEModel}
    I^{(\text{A})} \propto
    \sum_{i} Q^{(\text{A})}_{i}
    \sigma^{(\text{A})} \exp\left(- \, \frac{z_i}{\lambda\cos{\theta}}\right).
\end{equation}
Herein, \(i\) is the index of the layers and \(i = 0\) is at zero distance from the surface; \(z_i\geq 0\) is the depth to the top of the \(i\)-th layer; \(\lambda\) is the inelastic mean free path; \(\sigma^{(\text{A})}\) is the photoionization cross-section of specie \(\text{A}\); and \( Q^{(\text{A})}_{i}\) the quantity of \(\text{A}\) in layer \(i\), normalized such that \(\sum_i Q^{(\text{A})}_i = 1\). The attenuation length \(\lambda\) was assumed to depend exclusively on the kinetic energy ($E_{\mathrm{K}}$) of photoelectrons. Photoionization cross-sections \(\sigma^{(A)}\) were taken from Refs.~\onlinecite{Yeh1985atomic,yeh1993atomic}. With these assumptions, we could fit composition profiles to the relative intensities of the ARXPS signals obtained from the different species. The hyperparameters of the fit were the inelastic mean free path \(\lambda\) of the electrons, and the thickness \(\dd z\) of the model layers. Numerically, we normalized the measured and fitted intensities such that 
$\sum_{A'} I^{(A')}(\theta) = 1$ for every angle $\theta$ and performed the fit in this reduced space. 

For a given set of hyperparameters, we fitted the model depth profile to reproduce the measured relative intensities using a sequential quadratic programming method \cite{nocedal1999numerical}. Enforcing \(\sum_A Q^{(\text{A})}_i = 1\), we optimized within the constrained space without explicitly projecting any solutions onto the solution space. For the loss function, $L$, we used the weighted sum of squared residuals, i.e, a \(\chi^2\)-metric, with uncertainties/weights inherited from curve fitting of the {ARXPS} spectra:
\begin{equation}
    L(\underline{Q}) = \sum_{\theta}\sum_{A} \left(\frac{I^{(\text{A})}_{\text{model}}(\theta; \underline{Q}) - I^{(\text{A})}_{\text{measured}}(\theta)}{\delta I^{(\text{A})}}\right)^2,
\end{equation}
where, \(\underline{Q}\) encodes the composition of the layers, \(\theta\) the measured emission angles and \(\text{A}\) the chemical species. To ensure numerical stability, we employed a \emph{zero-th order regularization scheme}. 

A key limitation of this simple model is that it does not account for photoelectron diffraction effects \cite{Woodruff1994, Woodruff2007}. To mitigate this, we regularized the problem by introducing a LASSO-like \cite{hastie01statisticallearning} term, penalizing the amount of Sb in the layers:
\begin{equation}
    L_{\text{reg}}(\underline{Q}; \mu) = L(\underline{Q}) + \mu\sum_{j} Q^{(\text{Sb})}_{j},
\end{equation}
where \(\mu\) is a unitless regularization parameter and \(Q^{(\text{Sb})}_{j}\) the quantity of Sb in layer \(j\). The optimal choice for \(\mu\) depends on the hyperparameters of the model, i.e., the number of layers and their thickness \(\dd z\) relative to the inelastic mean free path \(\lambda\). For clarity, we have thus reported the absolute and relative increase of the unconstrained loss averaged over the number of data points -- $\chi_{(0)}^{2}$ and $\chi^{2}/\chi_{(0)}^{2}$, where $\chi^{2}$ is a constrained value.

When fitting hyperparameters, the conventional approach is to apply a cross-validation scheme. Our datasets have few enough data points to use a \emph{leave-one-out cross-validation approach}, although we found that using the entire dataset for training/testing gave virtually the same results. Since ${\mu}$ is not standardized, reporting nested cross-validation results quickly becomes unnecessarily complicated and opaque. In our case, the only recorded uncertainty is from fitting the core level spectra. Any variations in the angle-resolved data will come from stochastic noise, photoelectric diffraction, and Beer-Lambert-like attenuation. We expect that stochastic noise is a minor part of this, and thus we are not immediately risking overfitting the data. Another important point is that with weak random noise, the measured intensity should vary smoothly with the angle $\theta$. Said differently, the fit depends on the sampled angles which complicates a transparent cross-validation process.
\section{Stoichiometry of a Buried Alloy Layer}
\label{sec:BuriedLayer}

\begin{figure}
    \centering    \includegraphics{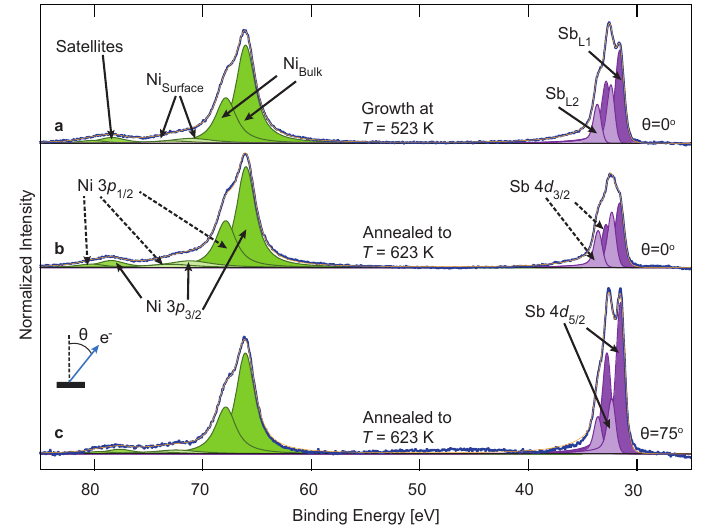}
    \caption{XPS ($h\nu=250$~eV) of the Ni~3\textit{p} and Sb~4\textit{d} core levels of NiSb. A larger amount of Sb was deposited onto heated Ni(111) to achieve more measurable Sb from the near-surface atomic layers (Table~\ref{tab:SamplePrepTC}, lower entry). The background has been subtracted and the spectra normalized to the area of the bulk $\mathrm{Ni}~3\mathit{p}$. \textbf{a}: After step-wise Sb deposition at $T=\SI{523}{\K}$. \textbf{b}-\textbf{c}: After post-growth annealing to $T=\SI{623}{\K}$. The angles $\theta$ relative to normal emission for all measured spectra have been indicated in the insets.}
    \label{fig:Supp_XPS}
\end{figure}

Our finite element model analysis has shown that exposing Ni(111) to Sb ad-atoms at elevated temperatures results in near-stoichiometric NiSb forming in the top atomic layers. Previously, Auger electron spectroscopy has suggested that besides supplying the activation energy necessary for alloying, annealing above room temperature will also facilitate Sb diffusing into the bulk Ni \cite{KRUPSKI2003}. Thus, to explore the growth of potentially thicker NiSb films on Ni(111), we deposited more Sb onto the heated substrate (for $\approx1$~hour), followed by a quick post-growth anneal to recrystallize the surface (additional details in Table~\ref{tab:SamplePrepTC}, second row). The resultant Ni~$3p$ and Sb~$4d$ core level spectra (Fig.~\ref{fig:Supp_XPS}) exhibit similar binding energies and spin-orbit splitting as their differently prepared counterparts. However, the absolute abundance of Sb is now larger, especially in terms of metallic Sb (L1). In relative terms, it is similarly abundant to the Sb alloying with the Ni (L2).

\begin{figure}[t]
    \centering
    \includegraphics[]{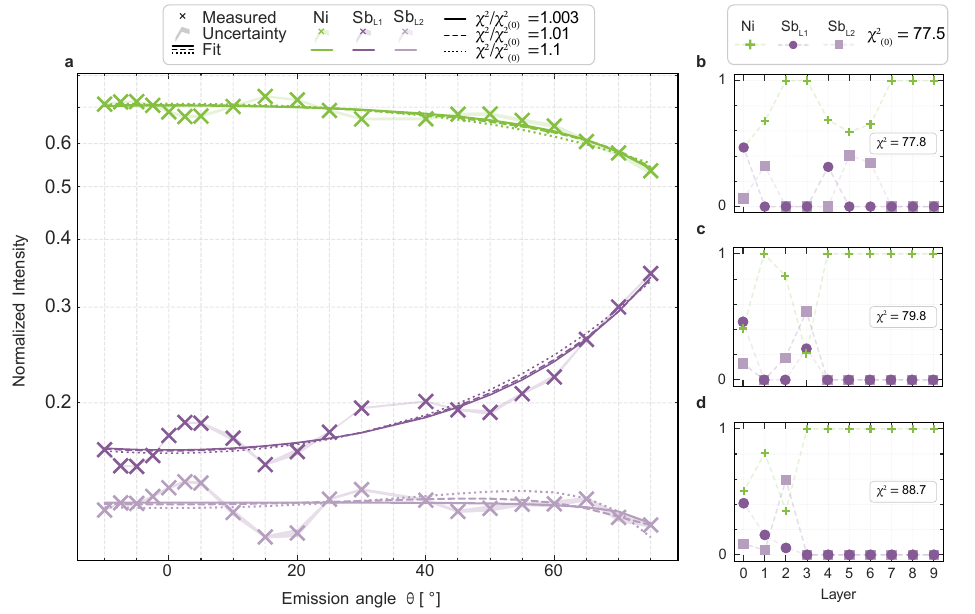}
    \caption{The Sb ad-atom distribution on and within Ni(111) during NiSb alloy formation. \\ \textbf{a}: Measured normalized intensities, $I$, of the different chemical species in the NiSb sample (Table~\ref{tab:SamplePrepTC}, lower entry), with uncertainties, as a function of emission angle \(\theta\). A finite element model with $3$ levels of regularization as indicated by the $\chi^{2}/\chi^{2}_{(0)}$ values (trivial, $\chi^{2}$ increases by 1\% and by 10\%, see details in Sec.~\ref{sec:lmodel}) was fitted to the measurements.
    \textbf{b}-\textbf{d}: The corresponding fitted depth profiles. The fit suggests a buried Sb layer even as we increase the regularization parameter and implicitly bias the fit towards placing Sb at the surface.    }
    \label{fig:SI_ARXPS}
\end{figure}

The spectra at normal ($\theta = 0^{\circ}$), and grazing ($\theta = 75^{\circ}$) emission in Fig.~\ref{fig:Supp_XPS} indicate that the additional metallic Sb is situated at the surface. This is further supported by more comprehensive ARXPS measurements and layer-by-layer analysis (Fig.~\ref{fig:SI_ARXPS}) using the same hyperparameters as before: a photoelectron attenuation length of $\lambda\approx\SI{0.5}{\nm}$; and $\dd z/\lambda=0.3$, giving \SI{0.15}{\nm} thick layers. The $\dd z/\lambda$ ratio was found by a grid search of the parameter space without a test/train split of the data. A cross-validation approach gave a virtually identical ratio. Details regarding the implementation can be found in Sec.~\ref{sec:lmodel}.

The unconstrained fit in Fig.~\ref{fig:SI_ARXPS}\textbf{b} indicates that an additional, narrow NiSb-like alloy region is located several atomic layers beneath the surface. To assess the robustness of this prediction, we implemented a {LASSO}-like \cite{hastie01statisticallearning} penalty on the amount of Sb to implicitly reduce its presence in the bulk Ni (details in Sec.~\ref{sec:lmodel}). Two additional and constrained model fits are shown in Figs.~\ref{fig:SI_ARXPS}\textbf{c} and \textbf{d}, where $\chi^{2}$ increased by 1~\% and 10~\%, respectively. Despite the penalty, a near-stoichiometric NiSb alloy profile persists beneath the surface. Our finite element model analysis thus indicates that thicker NiSb alloys can form on Ni(111) with sufficient Sb supply and dispersion. However, complementary investigations would be needed to verify the atomic arrangement and crystallinity of Sb in the subsurface layers.

%